\documentclass{aa}

\usepackage{graphicx}
\usepackage{txfonts}
\usepackage{color}
\usepackage{url}
\usepackage[flushleft]{threeparttable}
\usepackage[colorlinks=true, allcolors=blue]{hyperref}
\usepackage{orcidlink}
\usepackage{adjustbox}
\usepackage{multicol}
\usepackage{multirow}
\usepackage{xspace}
\newcommand {\gx}{{GX~349$+$2}\xspace}
\newcommand {\ixpe}{{IXPE}\xspace}
\newcommand {\nustar}{\textit{NuSTAR}\xspace}
\newcommand {\maxi}{{MAXI}\xspace}

\newcommand{\fluxcgs}{erg\,
s$^{-1}$\,cm$^{-2}$}
\newcommand{\lum}{erg\,s$^{-1}$}

\begin{document} 

\title{IXPE view of the Sco-like source \gx in the normal branch}

\titlerunning{IXPE view of the Sco-like source \gx in the normal branch}

\author{
Fabio La Monaca\inst{\ref{in:INAF-IAPS},\ref{in:UniRoma2}}\thanks{Corresponding author: fabio.lamonaca@inaf.it}\orcidlink{0000-0001-8916-4156}
\and 
Anna Bobrikova\inst{\ref{in:UTU}}\orcidlink{0009-0009-3183-9742}
\and 
Juri Poutanen\inst{\ref{in:UTU}}\orcidlink{0000-0002-0983-0049}
\and
Francesco {Coti~Zelati}\inst{\ref{in:ICE},\ref{in:IEEC},\ref{in:oabr_merate}}\orcidlink{0000-0001-7611-1581}
\and 
Maura Pilia\inst{\ref{in:Cagliari}}\orcidlink{0000-0001-7397-8091}
\and
Alexandra~Veledina\inst{\ref{in:UTU},\ref{in:Nordita}}\orcidlink{0000-0002-5767-7253}
\and 
Matteo Bachetti\inst{\ref{in:Cagliari}}\orcidlink{0000-0001-7397-8091}
\and
Vladislav Loktev\inst{\ref{in:UTU},\ref{in:Hel}}\orcidlink{0000-0001-6894-871X}
\and
Fei Xie\inst{\ref{in:Guangxi},\ref{in:INAF-IAPS}}\orcidlink{0000-0002-0105-5826}
}

\authorrunning{F. La Monaca et al.}

\institute{
        INAF--Istituto di Astrofisica e Planetologia Spaziali, Via del Fosso del Cavaliere 100, 00133 Roma, Italy \label{in:INAF-IAPS}
        \and
        Dipartimento di Fisica, Universit\`{a} degli Studi di Roma ``Tor Vergata'', Via della Ricerca Scientifica 1, 00133 Roma, Italy \label{in:UniRoma2} 
        \and 
        Department of Physics and Astronomy, 20014 University of Turku, Finland \label{in:UTU}
        \and
        Institute of Space Sciences (ICE, CSIC), Campus UAB, Carrer de Can Magrans s/n, 08193 Barcelona, Spain
        \label{in:ICE}
        \and
        Institut d'Estudis Espacials de Catalunya (IEEC), 08860 Castelldefels (Barcelona), Spain
        \label{in:IEEC}
        \and
        INAF--Osservatorio Astronomico di Brera, Via Bianchi 46, 23807 Merate (LC), Italy
        \label{in:oabr_merate}
        \and
        INAF--Osservatorio Astronomico di Cagliari, Via della Scienza 5, 09047 Selargius (CA), Italy
        \label{in:Cagliari}
        \and
        Nordita, KTH Royal Institute of Technology and Stockholm University, Hannes Alfv\'ens v\"ag 12, SE-10691 Stockholm, Sweden \label{in:Nordita}
        \and 
        Department of Physics, P.O. Box 64, 00014 University of Helsinki, Finland \label{in:Hel}
        \and
        Guangxi Key Laboratory for Relativistic Astrophysics, School of Physical Science and Technology, Guangxi University, Nanning 530004, China \label{in:Guangxi}}
        
\date{Received 23 May 2025, 09 July 2025}

\abstract{We present a detailed spectropolarimetric study of the Sco-like Z-source \gx, simultaneously observed with the Imaging X-ray Polarimetry Explorer (\ixpe) and Nuclear Spectroscopic Telescope Array (\nustar). During the observations \gx was found mainly in the normal branch. A model-independent polarimetric analysis yields a polarisation degree of $1.1\% \pm 0.3\%$ at a polarisation angle of $29\degr \pm 7\degr$ in the 2--8\,keV band, with ${\sim}4.1\sigma$ confidence level significance. No variability of polarisation in time and flux has been observed, while an energy-resolved analysis shows a complex dependence of polarisation on energy, as confirmed by a spectropolarimetric analysis. Spectral modeling reveals a dominant disc blackbody component and a Comptonising emitting region, with evidence of a broad iron line associated with a reflection component. Spectropolarimetric fits suggest differing polarisation properties for the disc and Comptonised components, slightly favouring a spreading layer geometry. The polarisation of the Comptonised component exceeds the theoretical expectations, but is in line with the results for other Z-sources with similar inclination. A study of the reflection's polarisation is also reported, with polarisation degree ranging around 10\% depending on the assumptions. Despite GX~349$+$2's classification as a Sco-like source, these polarimetric results align more closely with the Cyg-like system \mbox{GX~340$+$0} of similar inclination. This indicates that polarisation is governed primarily by accretion state and orbital inclination, rather than by the subclass to which the source belongs.}


\maketitle
%

\section{Introduction}\label{sec:intro}

Weakly magnetised neutron stars (WMNSs) in low-mass X-ray binaries (LMXBs) are among the brightest X-ray sources in the sky.  Without a strong magnetic field to channel the accretion flow, the strong gravity of the neutron star (NS) forces matter from a companion star to fall onto the NS, forming an accretion disc \citep{ShakuraSunyaev73} and relatively hot Comptonising medium near the surface of the NS, either in the form of the boundary layer \citep[BL; e.g.,][]{Shakura88,Popham01} or a spreading layer at the NS surface \citep[SL;][]{Lapidus85,inogamov1999}; the existence of this Comptonising medium situated between the inner disc and the NS surface is also suggested by the spectral and timing properties of quasi-periodic oscillations \citep{gilfanov2003, Revnivtsev06, Revnivtsev13}.
X-ray emission in WMNSs arises from these regions, producing a continuum spectrum with the disc emission slightly softer than that of the Comptonising media \citep{1984PASJ...36..741M}. Some sources show evidence for the presence of an extended accretion disc corona, a hot and optically thin plasma that spreads above and below the disc  \citep[see, e.g.,][]{White82, Parmar88, Miller00}; the presence of a reflected component \citep[see, e.g.,][for a review]{Ludlam24} and the scattering in the wind above the disc are sometimes also observed in these sources.

Based on the evolution of their luminosity and spectral hardness over time, WMNS are classified into Z and atoll sources \citep{hasinger89}. Z-sources are softer, brighter, and evolve more rapidly in time, shaping a Z-track in the Colour–Colour Diagram (CCD) or Hardness--Intensity Diagram (HID). This makes them perfect targets for the observations aimed at understanding the physics behind the evolution, geometry, and emission mechanisms of these systems.  The two most famous and brightest Z-sources, \mbox{Sco X-1} and \mbox{Cyg X-2}, show slightly different observational properties, and all the other Z-sources are sub-classified into Cyg-like or Sco-like sources. In particular, Cyg-like sources show all the traditional states of Z-sources: normal branch (NB); horizontal branch (HB); hard apex (HA) connecting HB with NB; flaring branch (FB); soft apex (SA) connecting NB and FB. In contrast, Sco-like sources are known to show less prominent HB, with a more significant change in intensity during flaring \citep{Church12}. 

Since 2021, WMNS-LMXBs have been actively observed by the Imaging X-ray Polarimetry Explorer \citep[\ixpe;][] {Weisskopf2022}. X-ray polarisation can be produced by electron scattering in the accretion disc atmosphere \citep{Chandrasekhar60, Dovciak08, Loktev22}, by the SL and/or BL \citep{st85,Farinelli24,Bobrikova24c}, by reflection of the central source radiation from the disc \citep{Lapidus85,Matt93,Poutanen96}, or by scattering off the wind \citep{Tomaru2024,Nitindala25}. 
Observations show a wide variety of phenomena, from an unexpectedly high polarisation in \mbox{4U~1820$-$303} \citep{DiMarco23b} to peculiar variability of the polarisation angle (PA) in \mbox{Cir~X-1} \citep{Rankin24} and \mbox{GX~13$+$1} \citep{Bobrikova24a, Bobrikova24b, DiMarco25b}. In Z-sources, a trend of dependence of polarimetric properties on the source state was observed in \mbox{GX~5$-$1} \citep{Fabiani24}, \mbox{XTE~J1701$-$462} \citep{Cocchi23}, and \mbox{GX~340$+$0} \citep{LaMonaca24gx340, LaMonaca25gx340NB}. In all three sources, the polarisation degree (PD) was higher in the HB states than in the NB/FB states. \mbox{Sco~X-1} \citep{LaMonaca24, LaMonaca25Sco} and \mbox{Cyg~X-2} \citep{Farinelli23} were observed in the SA and NB+FB, respectively, and showed low polarisation, consistent with the general trend, even if the X-ray PA of \mbox{Sco~X-1} was found to be neither aligned nor orthogonal to previous measurements of the direction of the radio jet. For reviews on \ixpe results on WMNSs, see \citet{Ursini24} and \citet{DiMarco25a}. 

\gx (also known as Sco~X-2) is a bright Sco-like Z-source. It is located at a distance of 9.2~kpc and has a luminosity that reaches 0.7--1.5 times the Eddington luminosity \citep{Grimm02}. \gx can delineate the entire evolutionary path on the HID in a day \citep[see, e.g.,][]{2004ApJ...600..358I, Coughenour18} and is known to have a strong, asymmetric and variable iron emission line \citep[see e.g.][]{2009ApJ...690.1847C, Cackett12}. The inclination of the source is constrained at 40\degr--47\degr\ by \cite{Iaria09} and at ${\sim}25\degr$ or ${\sim}35\degr$, depending on the spectral model, by \cite{Coughenour18}. 

\begin{table*}[!bth]
\centering
\caption{List of the simultaneous \ixpe and \nustar observations of \gx.}
\label{tab:exposure}
\begin{tabular}{llcccc}
\hline \hline
& Observation ID & Start (UTC) & Stop (UTC)  &Telescope& Exposure time (s)\\
 \hline\
\ixpe & 03003601 & 2024-09-06 22:52 & 2024-09-09 01:25 & DU 1 & 95583\\
 &  & & & DU 2 &  95849\\
 & & & & DU 3 & 95962\\
\hline
\nustar 1 & 91002333002 & 2024-09-07 05:51 & 2024-09-07 16:31 & FPMA & 7673\\
 & & & & FPMB & 7978\\
 \nustar 2 & 91002333004 & 2024-09-09 12:16 & 2024-09-09 00:36 & FPMA & 10058\\
 & & & & FPMB & 10426\\
\hline
\end{tabular}
\end{table*}

Here, we continue the investigation of the polarimetric properties of Z-sources and report a spectropolarimetric study of \gx. The results of the coordinated observations by \ixpe and \nustar, with a summary of the observations and data reduction, are presented in Sect.~\ref{sec:observations}. The results of spectral and polarimetric analyses are reported in Sect.~\ref{sec:results} and discussed in Sect.~\ref{sec:discussion}.

\begin{figure*}[!h]
   \centering
   \includegraphics[width=0.83\textwidth]{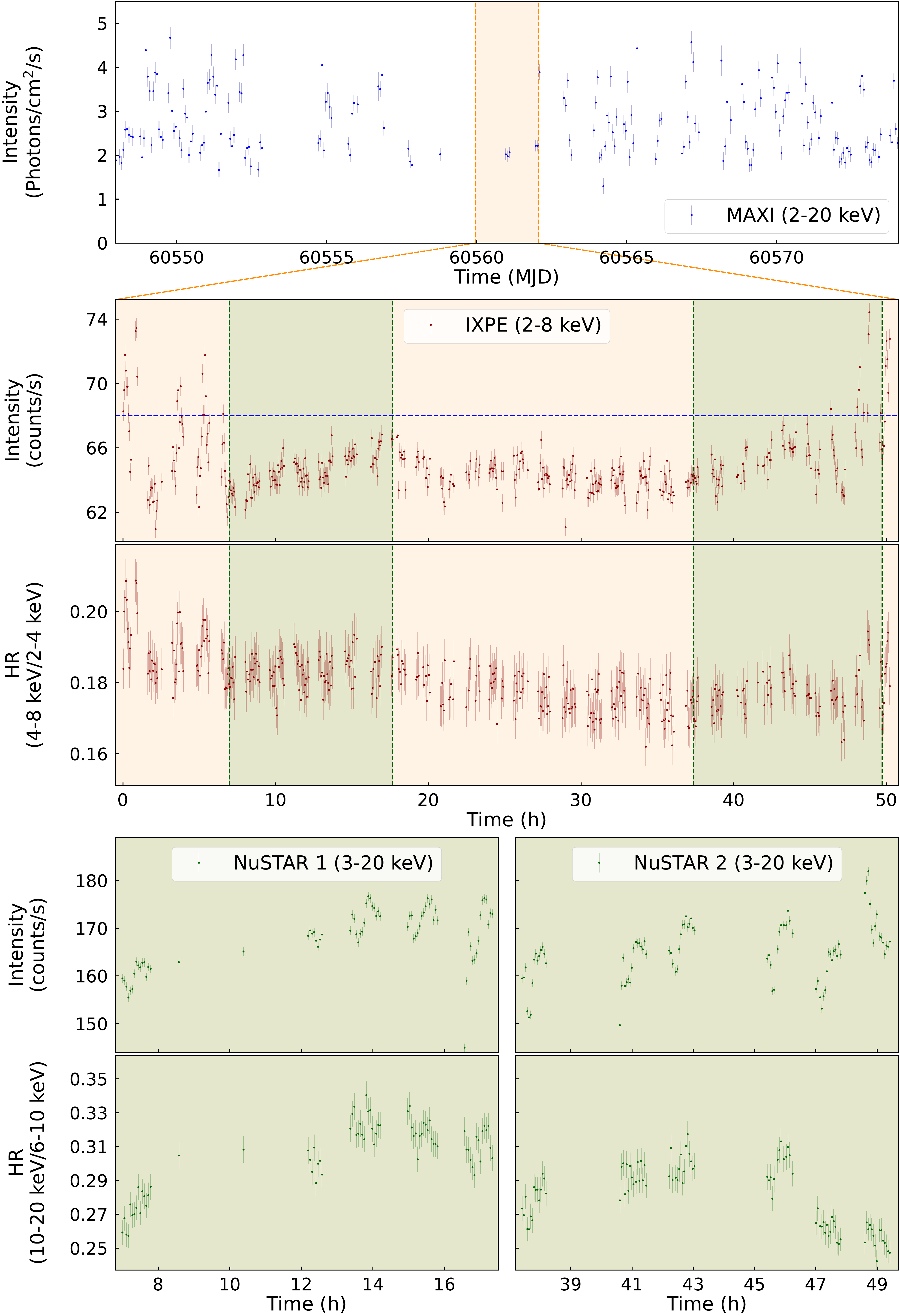}
   \caption{Light curves and HRs of \gx. \emph{Top panel}: \maxi light curve including the period of the \ixpe observation (orange shaded region). The \maxi data were binned in 1.5\,h intervals. \emph{Middle panel}: \ixpe light curve and HR binned in 200\,s intervals. The blue dotted line identifies the threshold for the high and low flux states used in the analysis.  The green shaded regions highlight the two \nustar simultaneous observations. \emph{Bottom panel}: \nustar light curves and HRs binned in 200\,s intervals. The \nustar count rate was obtained in the 3--20\,keV energy band. The time refers to hours since the start of the \ixpe observation. The observation IDs are reported in Table~\ref{tab:exposure}. 
   \label{fig:IXPE_NuSTAR_LC}}
\end{figure*}


\section{Observations and data reductions}\label{sec:observations}

The \ixpe and \nustar data from the observations, reported in Table~\ref{tab:exposure}, are publicly available at the NASA High-Energy Astrophysics Science Archive Research Center (HEASARC). X-ray data were extracted using standard pipelines and \textsc{ftools} included in \textsc{HEASoft} version 6.34 \citep{heasoft}. The light curves and HIDs are obtained using \textsc{stingray} \citep{stingray2,stingray1,Bachetti24}. Spectral and spectropolarimetric analyses were performed using the \textsc{xspec} X-ray spectral fitting package \citep{Arnaud96}. The \maxi \citep{MAXI} light curves are distributed publicly on the mission website.\footnote{\href{http://maxi.riken.jp/top/index.html}{http://maxi.riken.jp/top/index.html}}

\subsection{\ixpe}\label{subsec:ixpe}

\begin{figure}[!h]
   \centering
   \includegraphics[width=0.95\columnwidth]{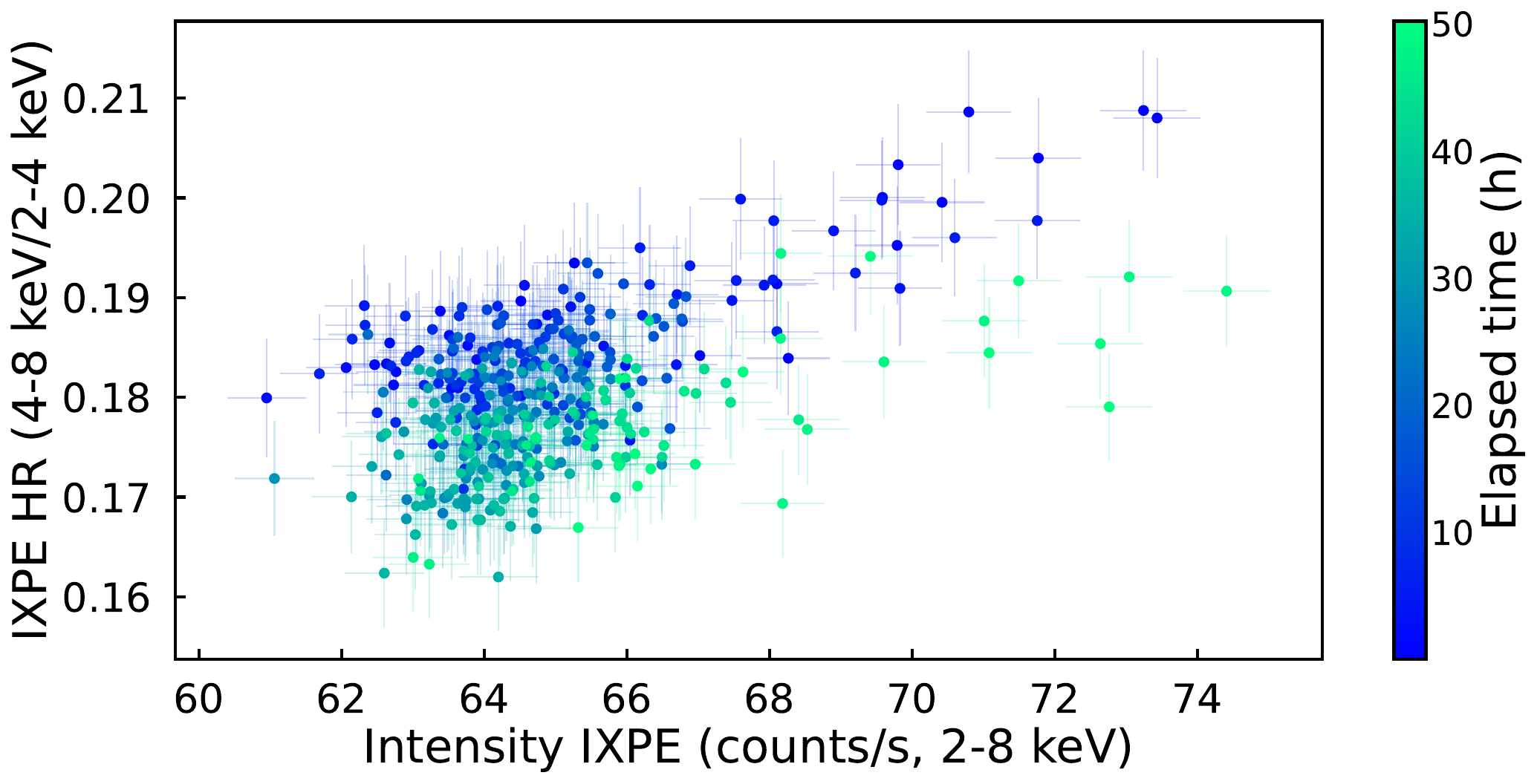}
   \caption{\ixpe Hardness-Intensity Diagram of \gx in 200-s time bins. The coloured points from blue to green report the elapsed time in hours since the start of the \ixpe observation. \label{fig:IXPE_HID}}
\end{figure}

\begin{figure}[!h]
   \centering
   \includegraphics[width=0.95\columnwidth]{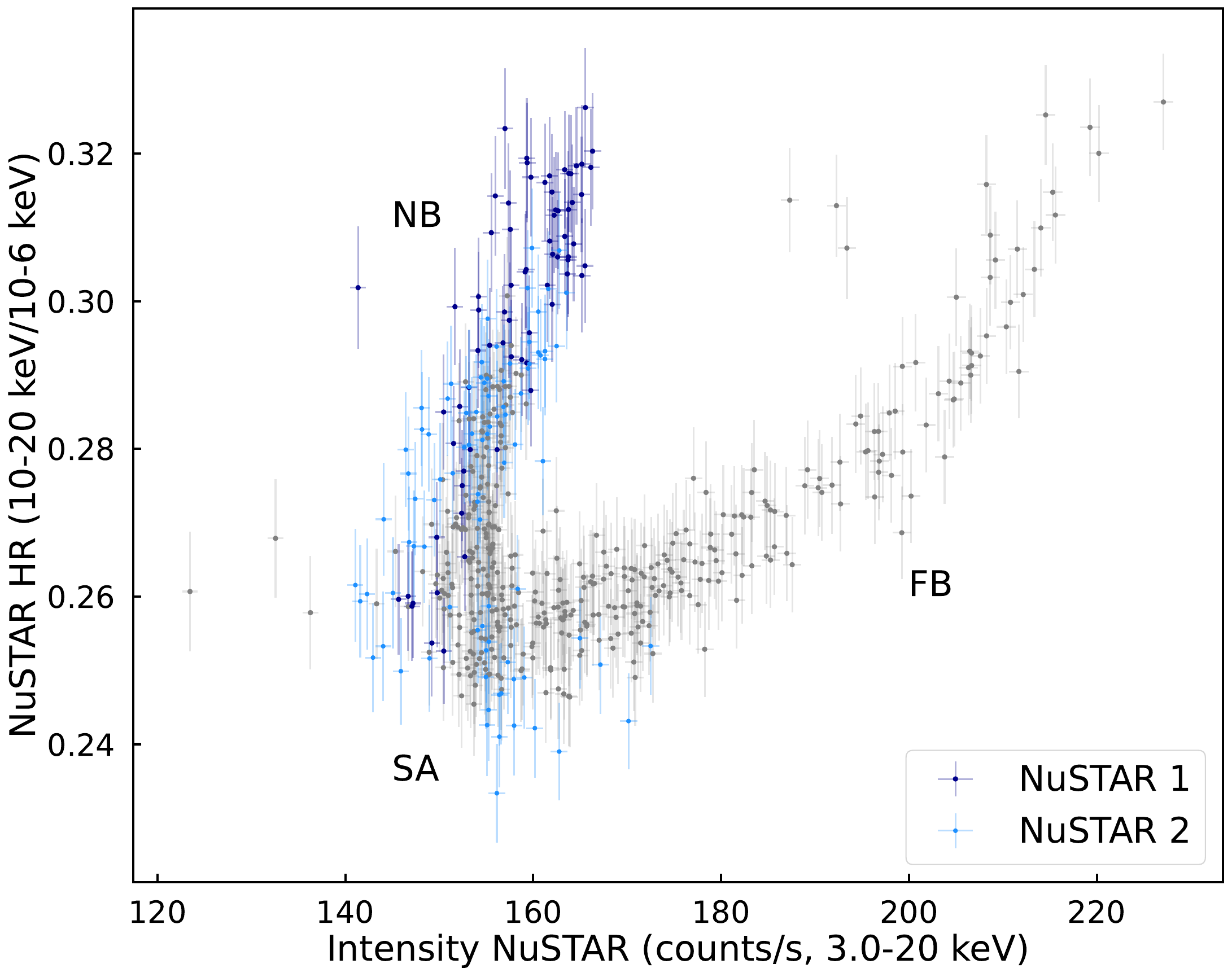} 
   \caption{\nustar Hardness-Intensity Diagram of \gx in 200-s time
bins. The dark and bright blue points represent the two \nustar observations reported in Table~\ref{tab:exposure} and Fig.~\ref{fig:IXPE_NuSTAR_LC} overlapping with archival \nustar observations (grey points). \label{fig:NuSTAR_HID}}
\end{figure}
\begin{figure}[!h]
   \centering
   \includegraphics[width=0.95\columnwidth]{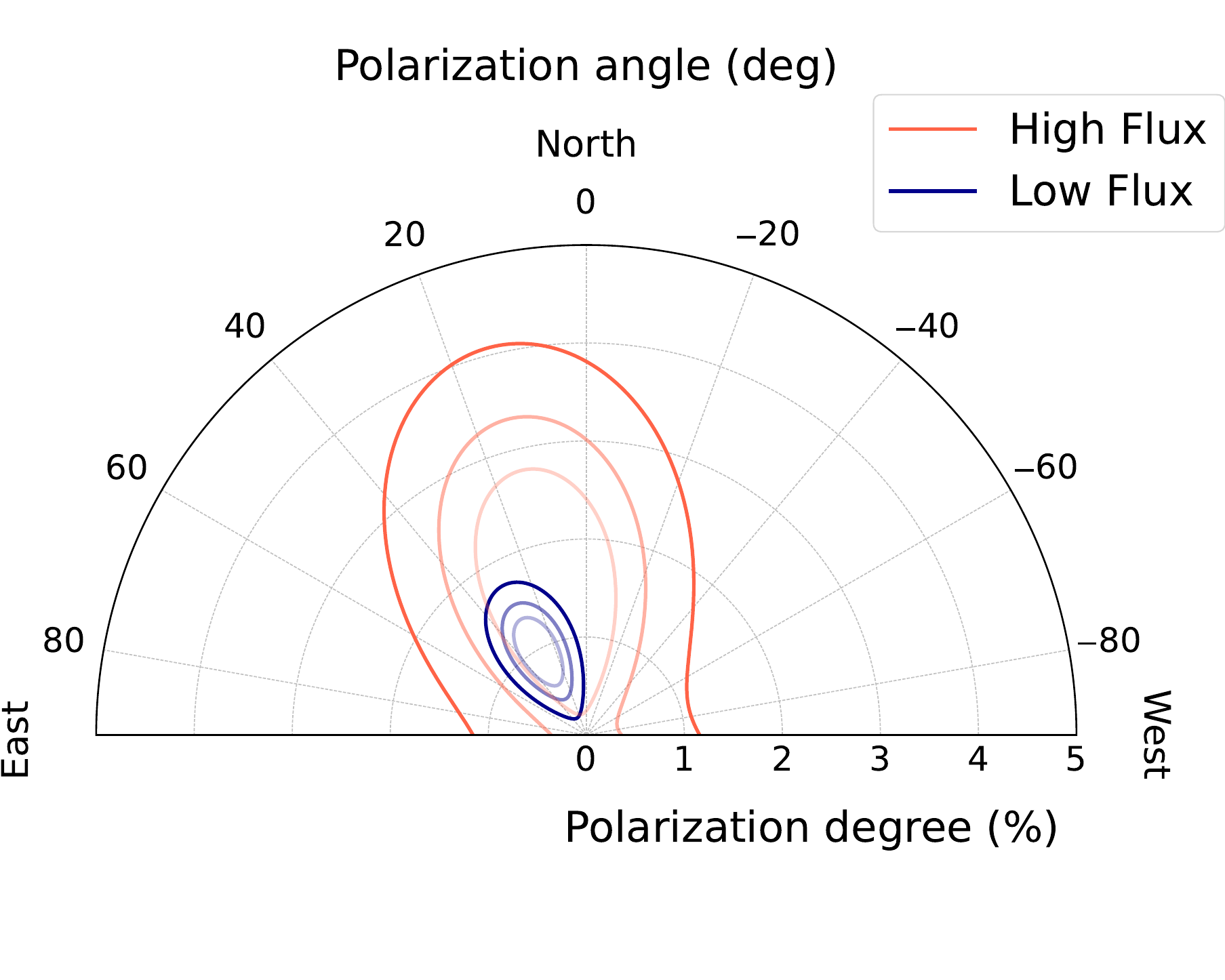} 
   \caption{Polar plot of the PD and PA for \gx in the 2--8 keV energy band obtained by \texttt{pcube} analysis when the \ixpe observation is divided into high and low flux states. The contours represent the allowed regions at 68\%, 90\% and 99\%~CL from the innermost to the outermost.\label{fig:contour_flux}}
\end{figure}

The NASA-ASI space telescope \ixpe was the first mission developed to investigate  X-ray linear polarisation. It was launched on 2021 December 9 and is equipped with three identical telescopes, each comprising a multi-mirror array and a polarisation-sensitive detector unit \citep[DU; see, e.g,][]{Baldini21, Soffitta2021, Weisskopf2022, DiMarco22b}.

\ixpe observed \gx from 2024 September 06 until 2024 September 09 for ${\sim}100$\,ks for each DU (see Table~\ref{tab:exposure}). The source region was selected using \textsc{SAOImageDS9}\footnote{\href{https://sites.google.com/cfa.harvard.edu/saoimageds9}{https://sites.google.com/cfa.harvard.edu/saoimageds9}} as a circular region centered on the source and with a radius of 100\arcsec. Background subtraction is not applied to the \ixpe data as suggested by \cite{DiMarco23a} in the case of bright sources. For the spectral and spectropolarimetric analysis, the \ixpe data were processed using the dedicated \ixpe software, \textsc{ixpeobssim} version 31.0.1 \citep{Baldini22}, with \ixpe CALDB response matrices 20240701 released on 2024 February 28. The model-independent analysis of polarisation, that is without any assumption on the spectral shape, was performed using the \textsc{ixpeobssim} \texttt{pcube} algorithm based on the \citet{Kislat15} approach. The \ixpe data were binned to have a minimum of 30 counts per bin for the Stokes $I$ spectrum, while a constant binning of 200 eV was applied to the Stokes $Q$ and $U$ spectra. The spectropolarimetric analyses reported in this paper were performed using the weighted approach \citep{DiMarco_2022}

\subsection{\nustar}\label{subsec:NuSTAR}
The Nuclear Spectroscopic Telescope Array (\nustar) is a space-based X-ray observatory that features two identical telescopes, named FPMA and FPMB, each of which comprises a multilayer coated Wolter-I grazing-incidence X-ray optic and a solid-state CdZnTe focal plane detector, enabling imaging and spectroscopy in the energy range of 3--79\,keV \citep{Harrison2013}.

\nustar observed the source twice during the \ixpe observation period of \gx. The corresponding observation IDs, the start and stop times, and the exposure times of each observation are reported in Table~\ref{tab:exposure}. In this work, the \nustar data extraction involved a multistep process performed using the standard pipeline by \nustar Data Analysis Software (NuSTARDAS), version 2.1.2 included in HEASoft 6.34, with the CALDB version 20241015. Considering the brightness of \gx, we included the statusexpr="STATUS==b0000xxx00xxxx000" keyword in NuPIPELINE to enhance data quality for sources with high count rates. For the correct estimation of the source and background emission, we extracted the source spectrum from a circular region centred on the source position with a radius of 120\arcsec, while the background spectrum was extracted from a source-free region with the same radius. The resulting spectra were binned to a minimum of 30 counts per bin.

\section{Data analysis and results}\label{sec:results}

Considering the variability of the Z-source, before performing any polarimetric analysis, we needed to identify the state of \gx during the \ixpe observation. Fig.~\ref{fig:IXPE_NuSTAR_LC} reports the long-term \maxi light curve, together with the simultaneous \ixpe and \nustar light curves. The hardness ratios (HRs) from \ixpe and \nustar as a function of time since the start of the \ixpe observation are also shown.
The \ixpe light curve shows stable behaviour except for the regions with higher count rates at the beginning and the end of the observation, although these regions do not show any evident corresponding hardness variation in the \ixpe band. Subsequently, we computed the HID using \ixpe and \nustar observations (see Fig.~\ref{fig:IXPE_HID} and Fig.~\ref{fig:NuSTAR_HID}, respectively) to identify the branch of the source at the Z-track. The HID of the two  \nustar observations,  compared with the HID obtained from the archival \nustar observations, shows that the source was in the NB during the first pointing and mainly in the NB with short periods in SA during the second and never entered deeply into the FB. Those two observations cover the first part and the end of the \ixpe observation; thus, we can conclude that \gx transitioned between NB and SA during the \ixpe observation with possibly short periods in FB.

\begin{table}[!h]
\centering
\caption{Energy-resolved polarisation for \gx.}
\label{tab:PD_PAvsEnergy}
\begin{tabular}{ccc}
\hline \hline
Energy Bin & PD & PA\\
(keV) & (\%) & (deg)\\
 \hline\
2.0--3.0 & $1.6 \pm 0.3$ & $18 \pm 6$ \\ 
3.0--4.0 & $0.4 \pm 0.3$ & $4 \pm 20$ \\ 
4.0--5.0 & $0.9 \pm 0.5$ & $53 \pm 15$ \\ 
5.0--6.0 & $1.0 \pm 0.7$ & $30 \pm 20$ \\ 
6.0--7.0 & $1.7 \pm 1.0$ & $56 \pm 18$ \\ 
7.0--8.0 & $4.0 \pm 1.9$ & $40 \pm 14$ \\ 
\hline
2.0--8.0 & $1.1 \pm 0.3$ & $29 \pm 7$ \\ 
\hline
\end{tabular}
\tablefoot{The polarisation is obtained with the \texttt{pcube} algorithm in the \ixpe nominal band and in 1\,keV-wide energy bins. The errors are reported at 68\%~CL.}
\end{table}

\begin{figure}[!h]
   \centering
   \includegraphics[width=\columnwidth]{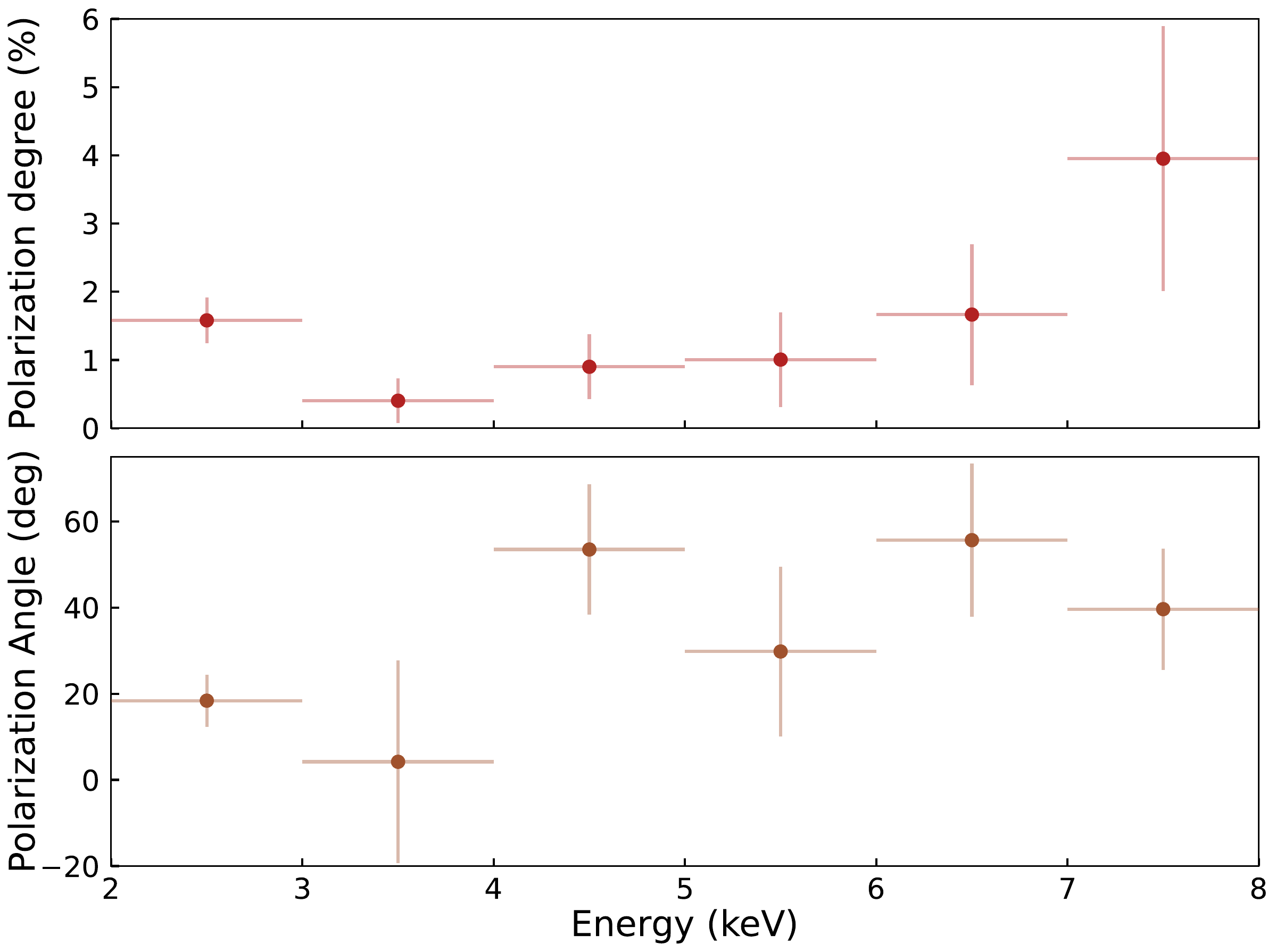}
   \caption{Energy-resolved polarimetric analysis of \gx: PD (\emph{top panel}) and PA (\emph{bottom panel}) in 1\,keV-wide energy bins. The values are also reported in Table~\ref{tab:PD_PAvsEnergy}. Errors are at 68\%~CL.\label{fig:PD_PAvsEnergy}}
\end{figure}

\begin{figure}
   \centering
   \includegraphics[width=\columnwidth]{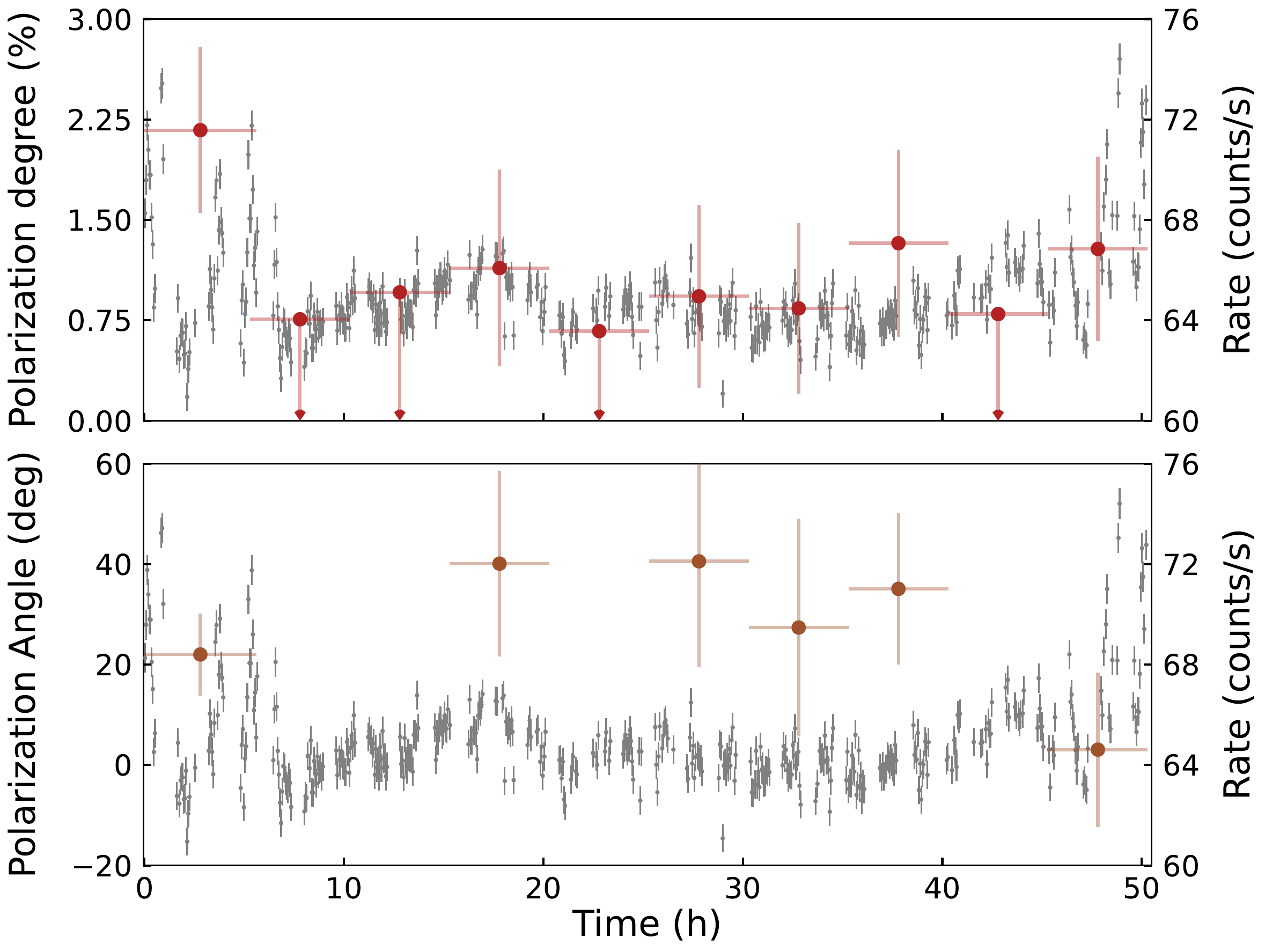}
   \caption{Time-resolved polarimetric analysis of \gx: PD (\emph{top panel}) and PA (\emph{bottom panel}). The \ixpe observation is divided into ten equal time bins of ${\sim}5.6$~hours. The errors and the upper limits are reported at 68\%~CL. \label{fig:PD_PAvsTime}}
\end{figure}

\subsection{Polarimetric analysis}

\citet{Kumar25} and \citet{Kashyap25}, which analysed the same \ixpe observation, tried to study separately the polarisation of NB, SA and FB. \citet{Kashyap25} selected only the part of the \ixpe observation that overlaps with the \nustar ones, reducing the analysed exposure time of \ixpe observation to only ${\sim}19\%$ of the total. This approach results in a measured polarisation below ${\sim}2\sigma$~confidence level (CL) in each branch with a loss of the major part of the \ixpe data. \cite{Kumar25} selected only the first part of the \ixpe observation as FB, even if they also report the source to be in FB at the end of the observation; this analysis corresponds to the first and last bin of the time-resolved analysis reported below in this section.
Here, we start the analysis trying to perform the same study, dividing the \ixpe observation into low and high flux states similarly with an approach applied in \cite{Rankin24}, \cite{LaMonaca25Sco} and \cite{DiMarco25b}. We set an intensity threshold at 68\,counts\,s$^{-1}$ on \ixpe data, see Fig.~\ref{fig:IXPE_NuSTAR_LC}. This threshold is obtained by considering the \ixpe intensity during the second \nustar observation. In fact, when \nustar\ showed that \gx is in the SA+FB (i.e., at the end of \nustar\ 2 observation), the \ixpe intensity is higher than 68\,counts\,s$^{-1}$. \ixpe light curve shows another period, before \nustar 1 observation, where intensity is higher than this value and with \ixpe HR variation; thus, we can confidently assume that the source was in the SA+FB, when the intensity is ${>}68$\,counts\,s$^{-1}$. We obtained the polar plot in Fig.~\ref{fig:contour_flux} showing compatibility within 68\%~CL between the low and high flux states in the 2--8\,keV energy band, using the \texttt{pcube} algorithm in \textsc{ixpeobssim} \citep{Baldini22}. 
Given this result and the very short period in FB+SA of \gx during the \ixpe observation, we performed the analysis using the whole duration of the \ixpe observation. The same approach was applied by \cite{Lavanya25}, which, using only IXPE data without any other simultaneous observations, concluded that the source was in NB+FB, and reported the polarisation of the whole dataset.

The model-independent analysis of the polarisation of the source in the 2--8\,keV energy band yields a PD of $1.1\%\pm0.3\%$ and a PA of $29\degr\pm7\degr$ with error at 68\%~CL. The significance of the measurement is $\sim$4.1$\sigma$~CL, and the probability of obtaining this polarisation in the case of an unpolarised source is $1.6\times 10^{-4}$. The results of the energy-resolved analysis of the polarisation are shown in Fig.~\ref{fig:PD_PAvsEnergy} and Table~\ref{tab:PD_PAvsEnergy}, where the polarisation is reported by dividing the \ixpe nominal energy band into 1\,keV-wide bins. It is interesting to note that the overall trend of PD with energy is similar to \mbox{GX~340$+$0} \citep{LaMonaca24gx340, LaMonaca25gx340NB}, hinting at an energy-dependent polarisation.

Some WMNSs showed evidence of time polarisation variabilities, such as \mbox{XTE~J1701$-$462} in the NB \citep{DiMarco25a} or \mbox{GX~13$+$1} \citep{Bobrikova24a, Bobrikova24b, DiMarco25b}. The polarisation properties were measured across ten evenly spaced time bins, each approximately 5.6 hours long. The results are consistent with a stable polarisation behaviour over time (see Fig.~\ref{fig:PD_PAvsTime}).
 
\subsection{Spectropolarimetric analysis}\label{subsec:spectropol}

\begin{figure}
\centering
\includegraphics[width=0.95\linewidth]{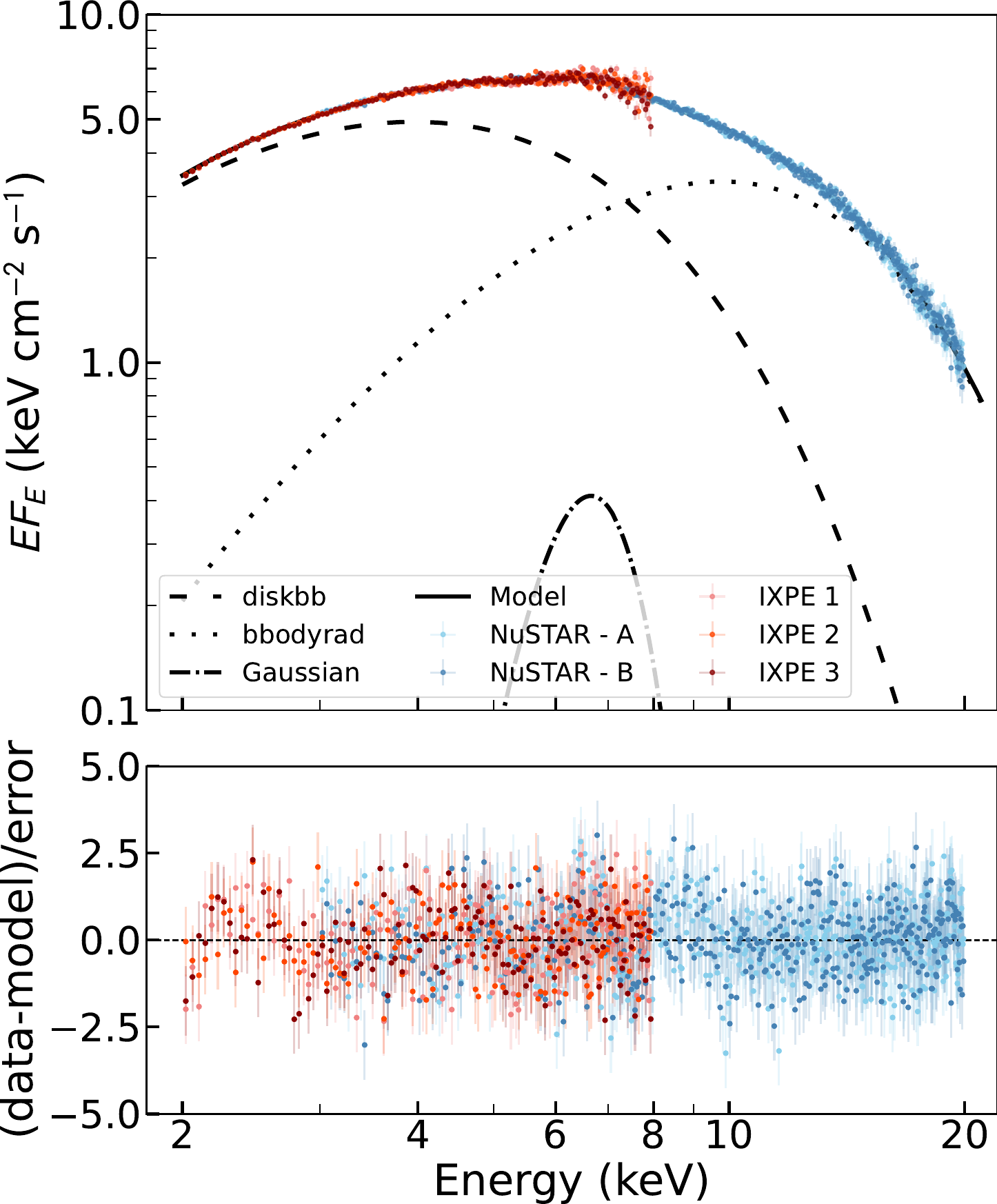}
\caption{Spectral energy distribution of \gx in $EF_E$ representation for the simultaneous fit using Model~A of \ixpe (red points) and \nustar (blue points) observations. The different spectral model components are reported in black lines for \texttt{diskbb} (dashed), \texttt{bbodyrad} (dotted), and \texttt{Gaussian}  (dot-dashed). The \emph{bottom panel} shows the residuals between the data and the best-fit Model~A. The best-fit parameters are reported in Table~\ref{tab:spectrum}.}
\label{fig:bestfitA}
\end{figure}

We performed a joint spectral analysis of \gx using contemporaneous observations from \ixpe and \nustar (we selected observation ID 91002333004 because it is more representative of the source behaviour during the \ixpe observation, as it also includes the interval with high flux). \nustar observation helps to constrain the broadband spectrum (3--20\,keV) due to its better spectral capabilities and extended spectral coverage compared to \ixpe. As is typical for this class of sources, we expect a spectrum characterised by a soft and a Comptonised component with possibly reflection features. Following the analysis of the spectrum of \gx in all different branches reported in \cite{Coughenour18}, we adopted a spectral model consisting of an absorbed multicoloured disc blackbody (\texttt{diskbb}) and a single-temperature blackbody (\texttt{bbodyrad}). The \texttt{diskbb} component describes the soft thermal emission from the geometrically thin, optically thick accretion disc, parameterised by the inner disc temperature and the normalisation related to the apparent inner radius, while the \texttt{bbodyrad} component accounts for the harder emission from a more compact, localised Comptonising region, likely associated with BL and/or SL, characterised by a blackbody temperature $kT_{\rm s}$ and a normalisation proportional to the square of the emitting radius $R_{\rm bb}$ of such a region. To achieve an acceptable fit for our data, we added a \texttt{Gaussian} line component to the continuum at ${\sim}6.4$\,keV to model the most prominent feature of the reflection spectrum. Thus, we fitted the data using the model \texttt{tbabs*(diskbb+bbodyrad+gauss)}. The \texttt{tbabs} component models the interstellar absorption, with abundances set according to \texttt{wilm} values \citep{2000ApJ...542..914W} and the hydrogen column density frozen at $N_{\rm H} = 0.5 \times 10^{22}$\,cm$^{-2}$ \citep{Dickey90, Kalberla05, HI4PI}. 
We tied all physical parameters between the \ixpe and \nustar datasets, allowing only cross-normalisation factors to vary. The best-fit parameters of this model and the reduced $\chi^2$ are reported in Table~\ref{tab:spectrum} as Model~A. The spectral energy distribution with residuals is shown in Fig.~\ref{fig:bestfitA}. The best-fit parameters and uncertainties were estimated from a Markov Chain Monte Carlo (MCMC) analysis as implemented in \textsc{xspec}. MCMC chains were initialised around the best-fit solution derived from standard minimisation $\chi^2$, using a chain length of $1\times10^5$ and 60 walkers, after a burn-in phase of $1\times10^4$ steps. 

\begin{figure} 
\centering
\includegraphics[width=0.95\linewidth]{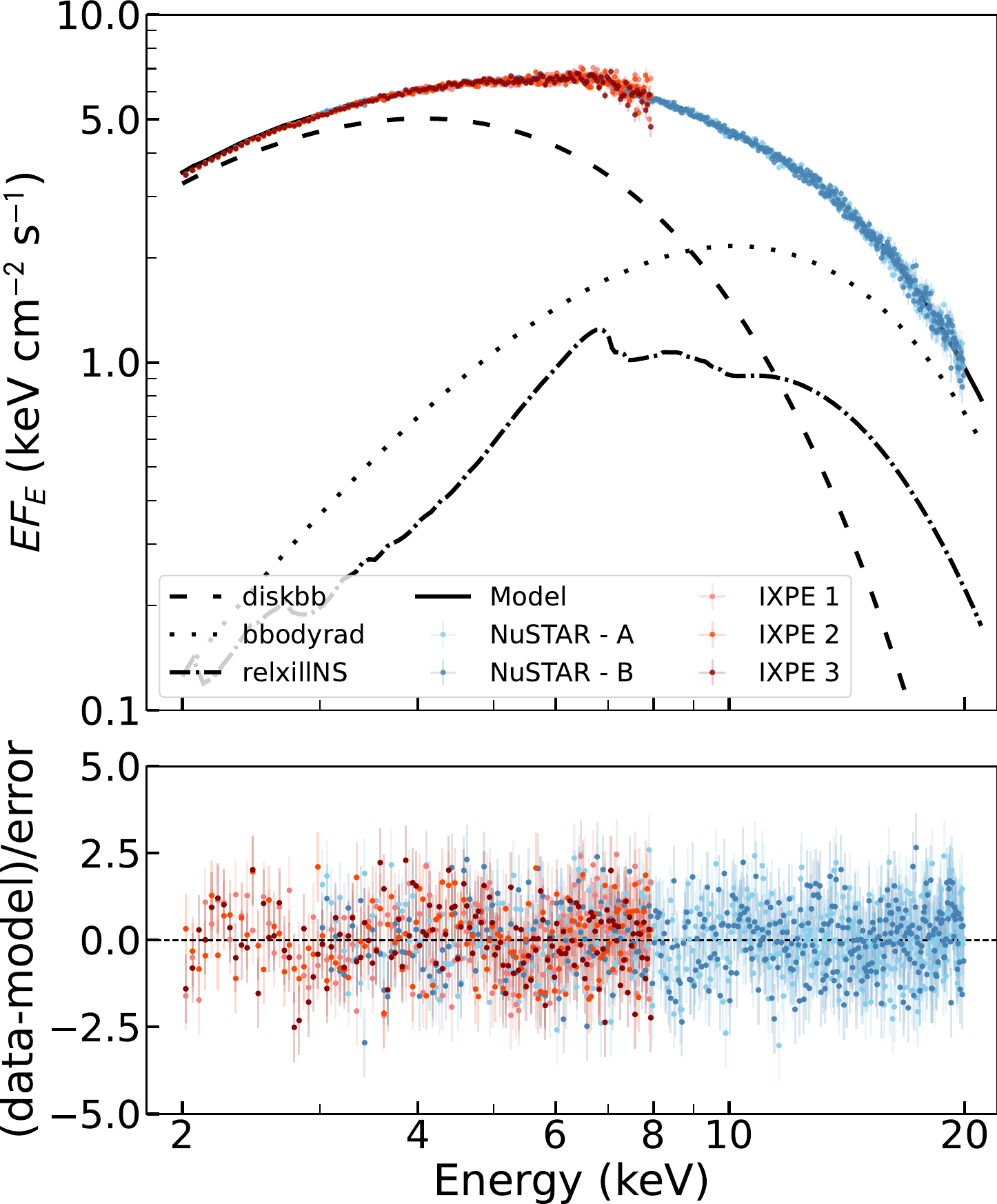}
\caption{Same as Fig.~\ref{fig:bestfitA} but for Model~B with the dot-dashed lines showing the \texttt{relxillNS}  component.} 
\label{fig:bestfitB}
\end{figure}

Given the presence of the Gaussian line and the potential Compton hump feature above 10\,keV, we extended our spectral model to include a reflection component using the \texttt{relxillNS} model, specifically developed to model the X-ray radiation reprocessed in the accretion disc around a NS \citep{Garcia22}. In this model, the incident spectrum was set to a blackbody with temperature equal to that of the \texttt{bbodyrad} component, therefore implying illumination from the BL and/or SL. The spin parameter $a$ was fixed to zero \citep{Galloway08, Miller_Miller16, Coughenour18, Ludlam24}. Moreover, the density parameter $\log N$ was fixed to 19, its maximum allowed value in this spectral model \citep{Garcia22}; the outer radius was set to $1000\,R_{\rm g}$; the reflection fraction parameter was set to $-1$ to select only the reflection spectral component; and the emissivity was set to $q=3$ as reported in \cite{2018MNRAS.475..748W} for an isotropic illuminating source as a belt, e.g., a BL or a SL. The best-fit parameters and uncertainties are reported in Table~\ref{tab:exposure} as Model~B and were estimated from MCMC with the same approach used for Model~A. The spectral energy distribution with residuals is shown in Fig.~\ref{fig:bestfitB}. We obtained an inclination of $32\degr \pm 1\degr$ (errors at 90\%~CL) in agreement with the previous results reported in \cite{Coughenour18} for the \texttt{relxillNS} model.

To investigate the model-dependent polarimetric properties of \gx and to search for any possible energy dependency of the polarisation, we performed a joint weighted spectropolarimetric fit of the \ixpe data using the best-fit spectral model obtained from the combined \ixpe and \nustar analysis. Thus, we started with the simplified Model~A fitting the \ixpe spectrum with all the parameters frozen at the values reported in Table~\ref{tab:spectrum}, and we applied the \textsc{xspec} \texttt{polconst}, \texttt{pollin}, and \texttt{polpow} polarimetric models. The \texttt{polpow} model gives unconstrained spectral indices, while \texttt{polconst} gives: $\textrm{PD}=0.7\%\pm0.3\%$ and $\textrm{PA}=32_{-14}^{+16}$\,deg with $\chi^2/\textrm{d.o.f.} = 783/737=1.06$ (errors at 90\%~CL).
When we adopt \texttt{pollin}, the results are $A_{1}=(-2.0_{-0.9}^{+0.4})\%$, $A_{\rm slope}=(0.8\pm0.3)$\%\,keV$^{-1}$, $\psi_{1}=(128_{-19}^{+17})$\degr, $\psi_{\rm slope}=(-14_{-5}^{+4}$)\degr\,keV$^{-1}$ with $\chi^2/\textrm{d.o.f.} = 773/735=1.05$ (errors at 90\%~CL).
The F-test gives F=4.8, which corresponds to an improvement probability of 99\%, showing a clear preference for the \texttt{pollin} model compared to \texttt{polconst}. 
Note that $A_1=-2$\% and $\psi_1=128\degr$ are equivalent to the PD of 2\% at PA=38\degr. 


Subsequently,  we adopted a separate \texttt{polconst} component to each spectral component, allowing the soft and Comptonised components to have their own polarisation. On the other hand, the \texttt{Gaussian} line component associated with the iron line is treated as unpolarised (by assigning its \texttt{polconst} amplitude to zero), as expected for a fluorescence line due to an isotropic process \citep[see, e.g.,][]{Churazov02, Veledina24}. The results of the weighted spectropolarimetric fit of the \ixpe Stokes $I$, $Q$ and $U$ spectra of \gx are reported in Table~\ref{tab:spectropol}. We obtained a $\chi^2/\textrm{d.o.f.} = 779/735=1.06$. 

We also performed weighted spectropolarimetric analysis using Model~B, aiming to estimate the contribution of the reflection component to the polarisation, whose continuum is expected to be highly polarised \citep{Matt93, Poutanen96}. Associating a \texttt{polconst} component to each spectral component of Model~B with the spectral parameters fixed at the best-fit values reported in Table~\ref{tab:spectropol} for this model, we obtained $\textrm{PD}=0.9\pm0.7$\% with $\textrm{PA}=5_{-23}^{+25}$\,deg at 90\%~CL for the \texttt{diskbb}, $\textrm{PD}<4\%$ at 90\%~CL for the \texttt{bbodyrad}, $\textrm{PD}<62\%$ at 90\%~CL for the \texttt{relxillNS} and the best-fit $\chi^2/\textrm{d.o.f.} = 769/733=1.05$ (see Model B1 in Table~\ref{tab:spectropol_rellxillNS}). This analysis confirms the difficulties in constraining the PD of the reflection component with respect to the other components, as already reported for other sources \citep[see, e.g.,][]{LaMonaca24, LaMonaca24gx340, LaMonaca25gx340NB}. When we fixed the PD values of the disc and Comptonised components to those obtained in Table~\ref{tab:spectropol} for Model~A, we obtained $\textrm{PD}<9\%$ for the \texttt{relxillNS} (see Model B2 in Table~\ref{tab:spectropol_rellxillNS}). Furthermore, if we set the PD of the Comptonised component to 0 (in \cite{st85} and \cite{Bobrikova24c} is expected $<1.5\%$) and fix the polarisation of the disc to the values reported in Table~\ref{tab:spectropol}, we obtained $\textrm{PD}=11\%\pm5\%$ with $\textrm{PA}=68_{-13}^{+12}$\,deg at 90\%~CL for the reflection component (see Model B3 in Table~\ref{tab:spectropol_rellxillNS}).

\begin{table*}[ht]
\centering
\caption{Best-fit parameters of the joint   \ixpe and \nustar spectral analysis for \gx.}
\label{tab:spectrum}
\begin{tabular}{lrcc}
\hline \hline
Model & Parameter (units) & Model~A & Model~B \\ \hline
\texttt{TBabs} & $N_{\rm H}$ ($10^{22}$ cm$^{-2}$) & [0.5] & [0.5] \\ \hline
\texttt{diskbb} & $kT_{\rm in}$ (keV) & $1.623\pm 0.010$ & $1.675 \pm 0.014$\\
                & norm ($[R_{\rm in}/D_{10}]^2\cos\theta$) & $108\pm2$  & $97\pm3$\\ 
                & $R_{\rm in}$ (km)$^a$& $10.4\pm0.2$& $9.8\pm0.3$\\
                \hline
\texttt{bbodyrad} & $kT_{\rm s}$ (keV) & $2.490_{-0.011}^{+0.010}$ & $2.586_{-0.019}^{+0.016}$\\
                & norm  ($[R_{\rm in}/D_{10}]^2$)& $ 17.5\pm0.5$ & $9.8_{-0.7}^{+0.9}$\\ 
                & $R_{\rm bb}$ (km)$^b$ & $3.85\pm0.11$ & $2.9_{-0.2}^{+0.3}$\\ \hline
\texttt{Gaussian} & $E$ (keV) & $6.41_{-0.05}^{+0.04}$ & --\\
                & $\sigma$ (keV) & $0.92_{-0.06}^{+0.08}$ & --\\
                & norm (photon~cm$^{-2}$~s$^{-1}$) & $0.022 \pm 0.002$ & --\\
                & Equivalent width (eV) & $ 0.147\pm 0.013$ & --\\ \hline
\texttt{relxillNS} & Emissivity & -- & [3.0] \\
                & $R_{\rm in}$ ({ISCO})$^c$ & -- & $1.8_{-0.3}^{+0.2}$ \\
                & $R_{\rm out}$ ($GM/c^2$) & -- & [1000] \\
                & Inclination (deg) & -- & $32 \pm 1$\\
                & $\log \xi$ & -- & $2.74_{-0.03}^{+0.04}$\\
                & $A_{\rm Fe}$ & -- & $0.77_{-0.06}^{+0.08}$\\
                & $\log N$ & -- & [19] \\
                & \text{norm} & -- & $0.0041\pm0.0004$\\
\hline
\multicolumn{2}{r}{$\chi^2/\textrm{d.o.f.}$} & 1371/1284 = 1.07 &  1299/1282 = 1.01\\ \hline
\multicolumn{4}{c}{Cross normalization factors} \\

& $C_{\rm \nustar-A}$ & [1.0] & [1.0]\\
& $C_{\rm \nustar-B}$ & $ 0.9839_{-0.0014}^{+0.0015}$ & $0.9838_{-0.0015}^{+0.0018}$ \\
&  $C_{\rm \ixpe-DU1}$ & $ 0.7264 \pm 0.0015 $ & $0.7241_{-0.0018}^{+0.0017}$\\
& $C_{\rm \ixpe-DU2}$ & $0.7303 \pm 0.0016$ & $0.7281_{-0.0015}^{+0.0018}$\\
& $C_{\rm \ixpe-DU3}$ & $0.7186 \pm 0.0016$ & $0.7165_{-0.0016}^{+0.0017}$\\ \hline
\multicolumn{4}{c}{Photon flux ratios in 2--8\,keV} \\
& $F_{\rm \texttt{diskbb}}/F_{\rm tot}$ & 0.74  & 0.76   \\
& $F_{\rm \texttt{bbodyrad}}/F_{\rm tot}$ &  0.24 &  0.15 \\
& $F_{\rm \texttt{gauss}, \texttt{relxillNS}}/F_{\rm tot}$ & 0.02  &  0.09 \\ \hline
\end{tabular}
\tablefoot{The estimated unabsorded flux in 2--8\,keV is $ 1.3\times10^{-8}$\,\fluxcgs, corresponding to a luminosity of ${\sim} 10^{38}$\,\lum\ for a distance to the source of 9.2\,kpc \citep{Grimm02}. The errors are reported at 90\%~CL.\\
$^{a,b}$  The distance of the source is assumed to be 9.2\,kpc \citep{Grimm02} and, for the \texttt{diskbb}, the inclination is assumed to be 32\degr\ as derived from the \texttt{relxillNS} model.\\
$^c$ The inner radius is given in units of innermost stable circular orbit (ISCO).
}
\end{table*}

\begin{table*}
\centering
\caption{Weighted spectropolarimetric analysis of \gx and \mbox{GX~340$+$0} using a spectral model with a soft disc component, a hard Comptonised one and a Gaussian Fe line.}
\label{tab:spectropol}
\begin{tabular}{llccc}
\hline \hline
Spectral component & &  \gx &  \mbox{GX~340$+$0} NB$^{a}$ &  \mbox{GX~340$+$0} HB$^{b}$\\
 \hline
\texttt{diskbb} & PD (\%) & $1.0 \pm 0.7$ &  <1.2  & $3.1 \pm 1.7$ \\ 
 & PA (\degr) & $6 _{-23}^{+24}$ & -- &  $-1 \pm 16$\\ 
 \hline
\texttt{Comptonisation}& PD (\%) &  $3.7 \pm 3.0$ & $4.3 \pm 1.8$ & $5.2 \pm 1.0$\\ 
 & PA (\degr) &$68 \pm 26 $ & $44 \pm 13$ & $44 \pm 5$\\
\hline
\end{tabular}
\tablefoot{A polarisation with zero amplitude is associated with the Gaussian Fe line component in all the spectra. The errors are reported at 90\%~CL.\\
$^{a}$ Weighted spectropolarimetric values in the NB for \mbox{GX~340$+$0} as reported by \cite{LaMonaca25gx340NB}.\\
$^{b}$ Weighted spectropolarimetric values in the HB for \mbox{GX~340$+$0} as reported by \cite{LaMonaca24gx340}.}
\end{table*}

\begin{table*}
\centering
\caption{Weighted spectropolarimetric analysis of \gx using Model~B of Table~\ref{tab:spectrum}.}
\label{tab:spectropol_rellxillNS}
\begin{tabular}{llccc}
\hline \hline
Spectral component & & Model B1 & Model B2 & Model B3\\
 \hline
\texttt{diskbb} & PD (\%) & $0.9\pm0.7$ &  [1.0]  & [1.0] \\ 
 & PA (\degr) & $5_{-23}^{+25}$ & [6] & [6]\\ 
 \hline
\texttt{bbodyrad}& PD (\%) &  <4 & [3.7] & [0]\\ 
 & PA (\degr) &-- & [68] & --\\
\hline
\texttt{relxillNS}& PD (\%) & <62 & <9 & $11 \pm 5$\\ 
 & PA (\degr) & -- & -- & $68_{-13}^{+12}$\\
 \hline
\end{tabular}
\tablefoot{The errors are reported at 90\%~CL.\\}
\end{table*}

\section{Discussion and conclusions}\label{sec:discussion}

\ixpe observed \gx for ${\sim}100$\,ks, see Table~\ref{tab:exposure}. During the \ixpe observation, two \nustar observations were carried out. The analysis of the \ixpe and \nustar light curves reported in Fig.~\ref{fig:IXPE_NuSTAR_LC} showed almost stable behaviour with two short periods with higher count rates without evident hardness variation in the \ixpe data. The \ixpe and \nustar HIDs, shown in Figs.~\ref{fig:IXPE_HID} and \ref{fig:NuSTAR_HID}, respectively, confirmed that, during the \ixpe observation, the source spent most of the time in the NB with short periods in SA and FB, mainly during the second \nustar observation toward the end of the \ixpe observation. In order to select possible FB periods,  we divided the \ixpe observation into two data sets with high and low flux; the result aligns with the previous ones for \mbox{GX 5$-$1} \citep{Fabiani24} and \mbox{Sco X-1} \citep{LaMonaca24, LaMonaca25Sco}, where differences in polarisation between NB, SA and FB are not observed. Moreover, time-resolved polarisation analysis, reported in Fig.~\ref{fig:PD_PAvsTime}, shows an almost constant behaviour with time, with all bins compatible at 68\%~CL, with the exception of the first time bin that is compatible with the average PD at $1.9\sigma$~CL. This first bin corresponds to the one identified as FB in \cite{Kumar25}, and the results are compatible with the findings of that paper. Noticeably, the source showed the same flux intensity also at the end of the observation, where the simultaneous \nustar one indicates short periods in SA+FB. Even if the first time bin seems to show higher PD, this is compatible within $1\sigma$~CL with the last time bin. Furthermore, as previously reported, when an analysis is performed selecting \ixpe data on the basis of the flux, the PD value is $1.5\%\pm0.8\%$, which is compatible within $1\sigma$~CL with both the first and the last time bin, but also with the value reported by \cite{Kumar25} for the FB. Considering that the source can show short time variability and rapid forward and backward transitions from NB to SA and FB, a flux-based selection criterion on the \ixpe data offers a more accurate approach to measuring FB polarisation than the method used by \cite{Kumar25}, based on a time selection that may include short intervals in which the source is in NB. This flux-resolved analysis displays compatibility of the PD between NB and FB (see Fig.~\ref{fig:contour_flux}) within 68\%~CL, while in \cite{Kumar25} a compatibility at $1.7\sigma$ is reported, but the latter estimate is obtained by using errors from \texttt{pcube} analysis, that can be slightly underestimated with respect to the confidence regions in the PD vs PA polar plane.

We performed a model-independent \texttt{pcube} polarimetric analysis of the entire observation and obtained a $\sim$4.1$\sigma$ detection of polarisation with $\textrm{PD}=1.1\%\pm0.3\%$ and $\textrm{PA}=29\degr\pm7\degr$ in the nominal 2--8\,keV energy band (errors at 68\%~CL). Those results are compatible with the ones reported by \citet{Lavanya25} for the whole observation. The measured polarisation for \gx is compatible with the results of other Z-sources observed by \ixpe in a similar state: \mbox{Cyg~X-2} in the NB \citep[$\textrm{PD}=1.8\% \pm 0.3\%$;][]{Farinelli23}, \mbox{GX~5$-$1} in the NB+FB \citep[$\textrm{PD}=1.8\%\pm0.4\%$;][]{Fabiani24}, \mbox{XTE~J1701$-$462} in the NB \citep[$\textrm{PD}<1.5\%$;][]{Cocchi23}, \mbox{GX~340+0} in the NB \citep[$\textrm{PD} = 1.4\%\pm0.3\%$;][]{LaMonaca25gx340NB}. For \mbox{Sco~X-1} in SA, \ixpe reported a $\textrm{PD}=1.0\% \pm 0.2\%$ at 90\%~CL. Moreover, \ixpe has shown that the polarisation of Z-sources changes with the source state and is higher (${\sim}4\%$) in the HB than in the NB \citep[see, e.g.,] []{Cocchi23, Fabiani24, LaMonaca24gx340, LaMonaca25gx340NB, DiMarco25a}. 

We fitted the simultaneous \nustar and \ixpe spectra with the purpose of modelling the soft and hard spectral components typically observed in the WMNSs spectra, often associated with the presence of reflection features. Following the approach reported in \cite{Coughenour18}, we modelled the continuum spectrum with \texttt{diskbb} to describe the soft disc emission and \texttt{bbodyrad} to describe the hard emission from BL and/or SL. The residuals reveal an excess above 6\,keV that can be associated with the presence of reflection from the inner disc. Thus, we added a broad Gaussian line to achieve an acceptable fit (see Model~A in Table~\ref{tab:spectrum} and Fig.~\ref{fig:bestfitA}). The spectrum of the observation is dominated by the soft component ($\simeq$74\% of the total flux).  We also tried to model the broadband spectrum by adding \texttt{relxillNS}, which provides a comprehensive physical description of the reflected emission tailored for NS \citep{Garcia22}. This model yielded an improved fit. The best-fit reflection parameters (see Model~B in Table~\ref{tab:spectrum} and Fig.~\ref{fig:bestfitB}) indicate moderate disc ionisation, iron abundance slightly below the solar one, and an inclination of ${\sim}32\degr$. The relatively low inclination obtained is in line with previous results \citep[see, e.g.,][]{Coughenour18}, and with the expectation for a Sco-like source \citep{Kuulkers96}. 

The energy-resolved model-independent analysis reported in Fig.~\ref{fig:PD_PAvsEnergy} hints at a complex dependence of PD and PA on energy. A similar trend is also present for \mbox{GX~340$+$0} \citep{LaMonaca24gx340, LaMonaca25gx340NB}. A hint of variation of the polarisation with the energy was also reported by \citet{Lavanya25} studying the polarisation in two energy bands. The best way to test this dependence is to perform a model-dependent analysis, reported here for the first time. The weighted spectropolarimetric analysis favours the \textsc{xspec} \texttt{pollin} model with respect to the \texttt{polconst} (see Sect.~\ref{subsec:spectropol}). The non-zero slopes for both PD and PA indicate a complex polarimetric behaviour that was already observed in other sources. For the Cyg-like \mbox{GX~5$-$1} variation with energy of only PA was reported \citep{Fabiani24}. By contrast, for \mbox{Sco~X-1},  both model-independent and spectropolarimetric analyses showed no evidence of variation of PD or PA \citep{LaMonaca24}. Energy-dependent polarisation has also been reported for some atoll sources. In \mbox{4U~1820$-$303}, a PD increase with energy and a PA rotation of 90\degr\ at 4\,keV were observed \citep{DiMarco23b}. In \mbox{GX~9$+$1}, \cite{Prakash25} reported a PD and PA variation at 4\,keV with an overall similar behaviour to the one reported for \mbox{GX~340$+$0}. In conclusion, a similar overall energy dependence polarisation appears to be common among several atoll and Z-sources.  

One possible explanation for the dependence of polarimetric properties on energy is the changing contribution of different spectral components. If the disc emission that dominates at lower energies is polarised differently from the BL and/or SL emission that is more prominent at higher energies, it would lead to a change in PA and PD with energy. Since the SL emission is expected to be polarised perpendicular to the disc plane \citep[with a deviation by up to 30\degr\ in particular geometries, see][]{Bobrikova24c}, and the disc emission is expected to be polarised in the disc plane \citep{Dovciak08, Loktev22}, we would expect a change of PA by ${\sim}60-90\degr$ between higher and lower energies. 
We performed a spectroscopic fit to estimate the contribution of each spectral component to the total emission and found that the spectrum is dominated by the disc emission, as reported in Table~\ref{tab:spectrum}, and only at the highest energies, above 7\,keV, the harder component dominates the spectrum, see Figs.~\ref{fig:bestfitA} and \ref{fig:bestfitB}. In that case, we cannot conclude the geometry of the accretion flow just from the energy-resolved analysis. To verify if the hard component originates from SL or BL, we fitted the disc and Comptonised components independently in a spectropolarimetric analysis. The results reported in Table~\ref{tab:spectropol} show a significant difference in PA between these components: $60\degr\pm35\degr$. The results favour an SL emission but cannot exclude the BL fully at 90\%~CL; thus, an SL geometry is preferred over a BL one. A similar conclusion was reported by \citet{Lavanya25}, although they noted that their study would benefit from a comprehensive broadband spectropolarimetric analysis, which is not presented, and that we provided in our analysis. The PD of the Comptonised emission exceeds the theoretical predictions \citep{Farinelli24, Bobrikova24c} but is in line with the results of other \ixpe observations of WMNSs, such as \mbox{GX~340$+$0}, which, even if it is classified as a Cyg-like source, has a similar inclination to \gx. The polarisation values of \mbox{GX~340$+$0} in the HB and NB are reported in Table~\ref{tab:spectropol} for comparison. The difference between PAs of the disc and Comptonised emission is ${\sim}60\degr$. This could be explained by a specific configuration of the SL, \citep[see Fig.~7 in][]{Bobrikova24c}, or it could be a hint towards a misalignment in the binary system. However, with the Comptonised component only dominating the energy spectrum above 7\,keV (i.e., near the edge of the \ixpe energy range), we conclude that it is impossible to constrain the geometric properties of the Comptonised emission uniquely. Another possibility to explain the higher polarisation may be the presence of an extended accretion disc corona as a further component of the emission of polarised light \citep{DiMarco25b, LaMonaca25gx340NB} or scattering of the source emission in the equatorial wind \citep{Nitindala25}, but no wind features were reported in the spectrum of this source \citep{Iaria09, Homan16, Coughenour18}.

We also investigated the contribution of the reflection to the polarisation by applying spectropolarimetric analysis with the \texttt{polconst} model to each spectral component of Model~B, see Table~\ref{tab:spectropol_rellxillNS}. This analysis confirmed the difficulties, also observed in other sources \citep[see, e.g.,][]{LaMonaca24, LaMonaca24gx340, LaMonaca25gx340NB}, in disentangling the polarisation of the Comptonised component from that of the reflection when they are all free to vary independently, resulting in an upper limit $\textrm{PD} < 63\%$ at 90\%~CL for the reflection component. Fixing the polarisation of the soft and hard components to the values reported in Table~\ref{tab:spectropol} for Model~A, the reflection PD becomes ${<}9\%$ at 90\%~CL. Moreover, if we assume the extreme scenario with an unpolarised Comptonisation component, the polarisation of the reflection component is $\textrm{PD}=11\%\pm5$. The value aligns with theoretical expectations,  which predict that the reflection PD could be up to $30\%$ depending on the inclination \citep{Matt93, Poutanen96}. A comprehensive polarimetric analysis of the reflection component using a self‑consistent reflection model was not performed by \cite{Kashyap25}, \cite{Kumar25} and \cite{Lavanya25}. Furthermore, \cite{Kumar25} ascribed the polarisation of the reflection to the Gaussian line, even though such a line is expected to be almost unpolarised \citep{Churazov02, Veledina24} and the polarisation of the reflection arises mostly from the continuum \citep{Matt93, Poutanen96}.  

Although \gx is formally classified as a Sco-like source, its polarisation properties presented in this study -- specifically the behaviour of PD and PA  with energy and the comparable PD values of the Comptonised component -- align more closely with those observed in the Cyg-like source at similar inclination \mbox{GX~340$+$0}, rather than with \mbox{Sco~X-1}. These findings confirm that the Sco-like and Cyg-like classifications do not reliably predict polarisation behaviour. Instead, the polarisation is driven mainly by the accretion state (i.e., the position of the source along the Z track) and the orbital inclination of the system. 

\begin{acknowledgements}
This research used data products provided by the IXPE Team (MSFC, SSDC, INAF, and INFN) and distributed with additional software tools by the High-Energy Astrophysics Science Archive Research Center (HEASARC), at NASA Goddard Space Flight Center (GSFC). The Imaging X-ray Polarimetry Explorer (IXPE) is a joint US and Italian mission.  This research has made use of the MAXI data provided by RIKEN, JAXA and the MAXI team.
The authors acknowledge the NuSTAR team for scheduling the observations. In particular, the authors acknowledge Karl Forster, Daniel Stern, and Fiona A. Harrison. 
FLM is supported by the Italian Space Agency (Agenzia Spaziale Italiana, ASI) through contract ASI-INAF-2022-19-HH.0, by the Istituto Nazionale di Astrofisica (INAF) in Italy, and partially by MAECI with grant CN24GR08 “GRBAXP: Guangxi-Rome Bilateral Agreement for X-ray Polarimetry in Astrophysics”.
AB is supported by the Finnish Cultural Foundation grant No. 00240328.
AV acknowledges the Academy of Finland grant 355672. Nordita is supported in part by NordForsk.
FX is supported by National Key R\&D Program of China (grant No. 2023YFE0117200), and National Natural Science Foundation of China (grant No. 12373041 and No. 12422306), and Bagui Scholars Program (XF).
\end{acknowledgements}

%
%
\bibliographystyle{yahapj}
\bibliography{biblio}
%
\label{LastPage} 
\end{document}